\documentclass[journal=jacsat,manuscript=article]{achemso}

\usepackage[version=3]{mhchem} % Formula subscripts using \ce{}
\usepackage[T1]{fontenc}
\usepackage[latin9]{inputenc}
\usepackage{xcolor}
\usepackage{amsmath}
\usepackage{amssymb}
\usepackage{graphicx}
\usepackage{comment}
\usepackage{siunitx}
\usepackage{textcomp}
\usepackage[nomarkers,figuresonly]{endfloat} %for igures
\RequirePackage[a-1b]{pdfx}

\usepackage[normalem]{ulem}

\makeatletter
\usepackage{siunitx}

\makeatother
\usepackage{babel}
\author{Xiaofei Xiao}
\email{xiaofei.xiao15@imperial.ac.uk}
\affiliation{The Blackett Laboratory, Imperial College London, London SW7 2AZ, United Kingdom}

\author{Stefan A. Maier}
\affiliation{The Blackett Laboratory, Imperial College London, London SW7 2AZ, United Kingdom}
\alsoaffiliation{Chair in Hybrid Nanosystems, Nanoinstitute M\"{u}nchen, Faculty of Physics, Ludwig-Maximilians-Universit\"{a}t M\"{u}nchen, 80539 M\"{u}nchen, Germany}

\author{Vincenzo Giannini}
\affiliation{Instituto de Estructura de la Materia (IEM-CSIC), Consejo Superior de Investigaciones Cient\'{i}ficas, Serrano 121, 28006 Madrid, Spain}

\title{Ultrabroad-Band Direct Digital Refractive Index Imaging Based on Suspended Graphene Plasmon Cavities}

\abbreviations{IR,NMR,UV}
\begin{document}

\begin{tocentry}

\end{tocentry}


\begin{abstract}
Mid-infrared spectroscopy is essential for chemical identification and compositional analysis, due to the existence of characteristic molecular absorption fingerprints. However, it is very challenging to determine the refractive index of an analyte at low concentrations using current photonic systems in a broad mid-infrared spectral range. We propose an imaging-based nanophotonic technique for refractive index determination. The technique is based on deeply subwavelength graphene plasmon cavities and allows for the retrieval of molecular concentration. This method features a two-dimensional array of suspended graphene plasmon cavities, in which the extremely high field enhancement and extraordinary compression of graphene plasmons can be realized simultaneously by combining shallow and deep cavities. This enables resonant unit cells to be read out in the spatial absorption pattern of the array at multiple spectral points, and the resulting information is then translated into the refractive index of the analytes. The proposed technique gives complementary information compared with the current nanophotonic techniques based on molecular absorption, including the ability of the refractive index measurement, the ultra-broadband measurable spectral range, and the small volume of the analyte required, thereby pushing the potential for refractometric sensing technologies into the infrared frequencies.
\end{abstract}

%\date{\today}
%\maketitle


\newpage
\section{Introduction}

\begin{figure*}[htbp]
%\begin{centering}
\includegraphics[width=1\linewidth]{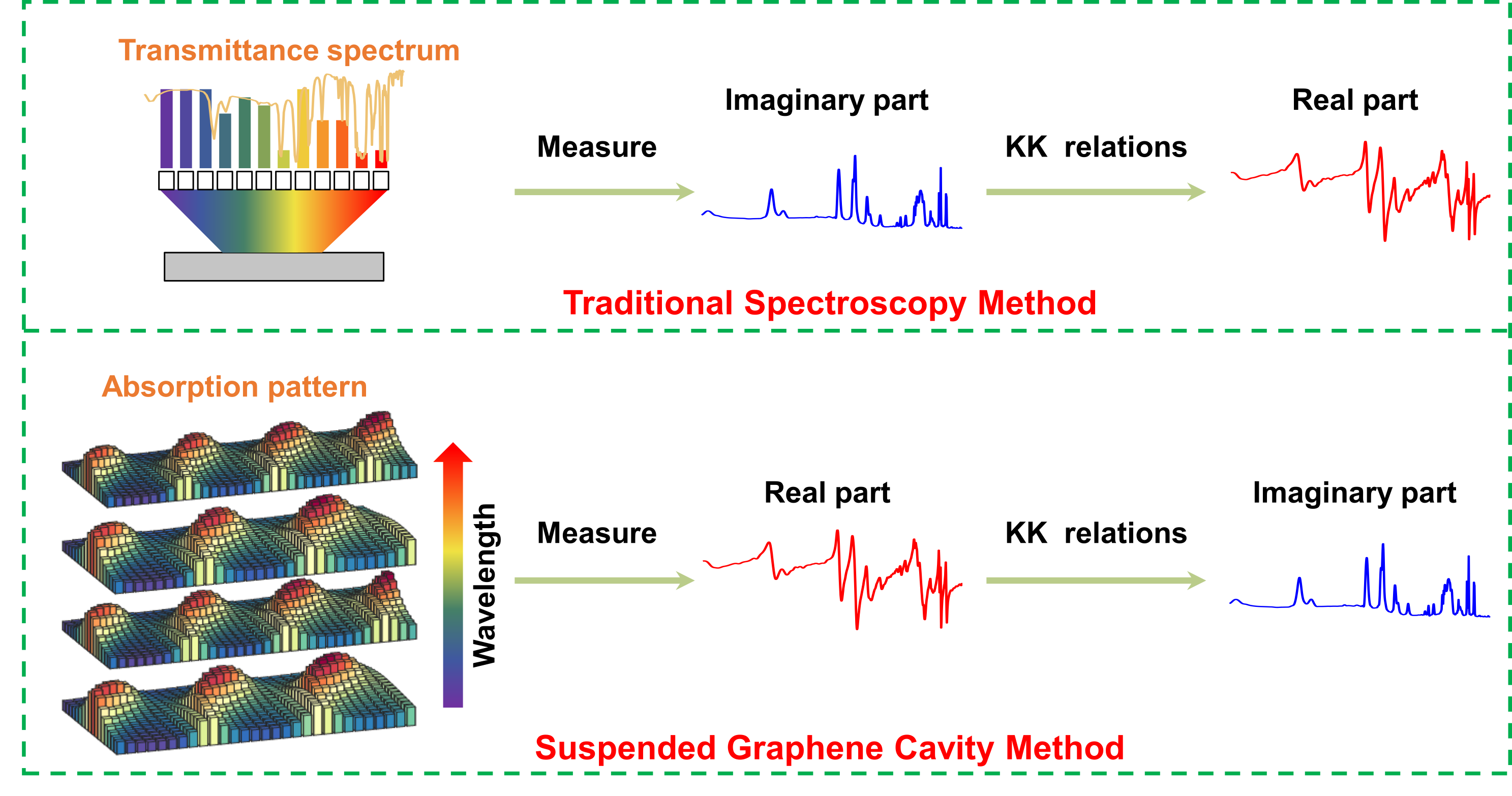}
%\par\end{centering}
\caption{\label{fig:comparassion} Different operating mechanisms that enable retrieving the complex refractive index of an analyte. Within the traditional bulk method, using a spectrometer (top path), the transmittance spectrum is measured. Based on Beer-Lambert's relation the imaginary part of the refractive index can be calculated. This method is based on the measurement of the absorption of the analyte, and thus the imaginary part of the refractive index of the molecules. An alternative approach (bottom path), the proposed suspended graphene cavity (SGC) method, is to measure the real part of the refractive index, which can be done by measuring the resonance pattern of a sensor array. Resonance, resulting in high absorption, occurs in some structures, only when resonance conditions are matched. The information hidden in the absorption pattern can be used to extract the refractive index of the detected molecules. KK relations are short for Kramers-Kronig relations.}
\end{figure*}

%\textbf{\textcolor{red}{Significance of mid-infrared spectra}}
The mid-infrared (MIR) spectrum is known for the presence of characteristic molecular absorption fingerprints originating from the intrinsic vibrational modes of chemical bonds, and thus the MIR spectrum is critical for chemical identification and structural characterization \cite{chalmers2002handbook}. MIR optical sensing technologies, such as MIR absorption spectroscopy, allow for the direct characterization of molecular structures, and have been recognized as powerful, non-destructive, label-free techniques for chemical analysis \cite{mizaikoff2013waveguide}. Conventionally, spectral analysis is achieved using macro systems, such as Fourier-transform infrared (FTIR) spectrometers, which measure the transmittance or emission spectrum of the analyte using gratings \cite{Rivas2008}. The analyte can be characterised based on either the retrieved refractive index (Fig. \ref{fig:comparassion}) or the existence of the characteristic molecular absorption fingerprints. However, this bulk approach usually requires complex and expensive equipment, such as FTIR spectrometers, and suffers from low sensitivity when detecting signals from small volumes of samples, because of the mismatch between MIR wavelengths ($\sim $ \textmu m) and the dimensions of molecules ($\sim$ nm).

 %\textbf{\textcolor{red}{Current nanophononics systems}}
 %The refractive index of a matter is an important parameter and it exhibits theoptical properties of the material.
To tackle this problem, nanophotonic systems with strong near-field enhancement of subwavelength resonators have been explored in MIR optical sensing. Recent sensor-on-chip approaches based on the surface-enhanced infrared absorption (SEIRA) have been realized based on various nanophotonic platforms \cite{li2017infrared,leitis2019angle,tittl2018imaging,neubrech2017surface,dong2017nanogapped,singh2016spr,chalmers2002handbook}, which potentially allow for a simplified and inexpensive sensor design that is well suited to miniaturization. However, the achieved performance is still far from ideal. Most platforms are based on near-field enhancement of subwavelength resonators; however, the tuning band range of the resonance frequency in current platforms only covers a small spectral region. Furthermore, although plasmonic or all-dielectric resonators are used, the field enhancement ($|E/E_0|^2$) is constrained to (usually) less than $10^3$ and field compression is limited to a scale around $\lambda_0/10$ (where $\lambda_0$ is the free-space wavelength) \cite{Giannini2008a,Giannini2011,Biagioni,Jahani2016}. This diminishes the sensitivity when looking at nanometre-scale samples at mid-IR wavelengths. In fact, most current sensor-on-chip approaches are based on the detection of characteristic molecular absorption \cite{singh2016spr,tittl2018imaging}. Thus, they  can be used for differentiating types of molecules; however, using these techniques it is challenging to obtain a quantitative characterization of properties of the analyte such as the refractive index. This is due to the low signal-to-noise ratio originating from the relatively small imaginary part of the refractive indices of molecules and the small volume of the analyte used in these approaches.

 %\textbf{\textcolor{red}{MIR refractive index sensor using graphene}}
 Within a Fabry-P\'{e}rot cavity, light reflects many times and forms a standing wave when resonance conditions are matched. The real part of the refractive index of the cavity materials determines the occurrence of the resonance, which is visible in the absorption of the system.  Therefore, the visibility of the resonance can be used to measure the refractive index of the material in the cavity. Such a technique has been realized in bulk systems \cite{ran2008laser,ran2009miniature}. However, it is difficult to realise in a deeply subwavelength system, due to the limited field compression and/or the large loss. Graphene plasmons \cite{grigorenko2012graphene,garcia2014graphene,rappoport2020understanding,wang2019graphene} (GPs) can overcome this limitation by exploiting their extremely short wavelength, strong field confinement, and large field enhancement. Furthermore, GPs have a frequency range covering the mid- and far-IR and terahertz bands \cite{low2014graphene}, and can be dynamically tuned by changing the Fermi energy via an external gate voltage \cite{fei2012gate,chen2012optical,rufangura2020towards}. This strong tunability combined with a broad spectral reach make GPs highly promising for various vital applications such as modulators \cite{low2014graphene} and photodetectors \cite{koppens2014photodetectors,shautsova2018plasmon,konstantatos2018current,nguyen2020ultra}. In the last decade, many tunable sensors \cite{ogawa2020graphene,epstein2020far}, including mid-infrared sensors \cite{rodrigo2015mid,xiao2016graphene,wenger2017high,yi2019dual}, based on graphene have been proposed and investigated, indicating that graphene could offer a great opportunity to significantly improve the performance of devices used for refractometric detection in the infrared band.
 
\begin{figure*}[htbp]
%\begin{centering}
\includegraphics[width=1\linewidth]{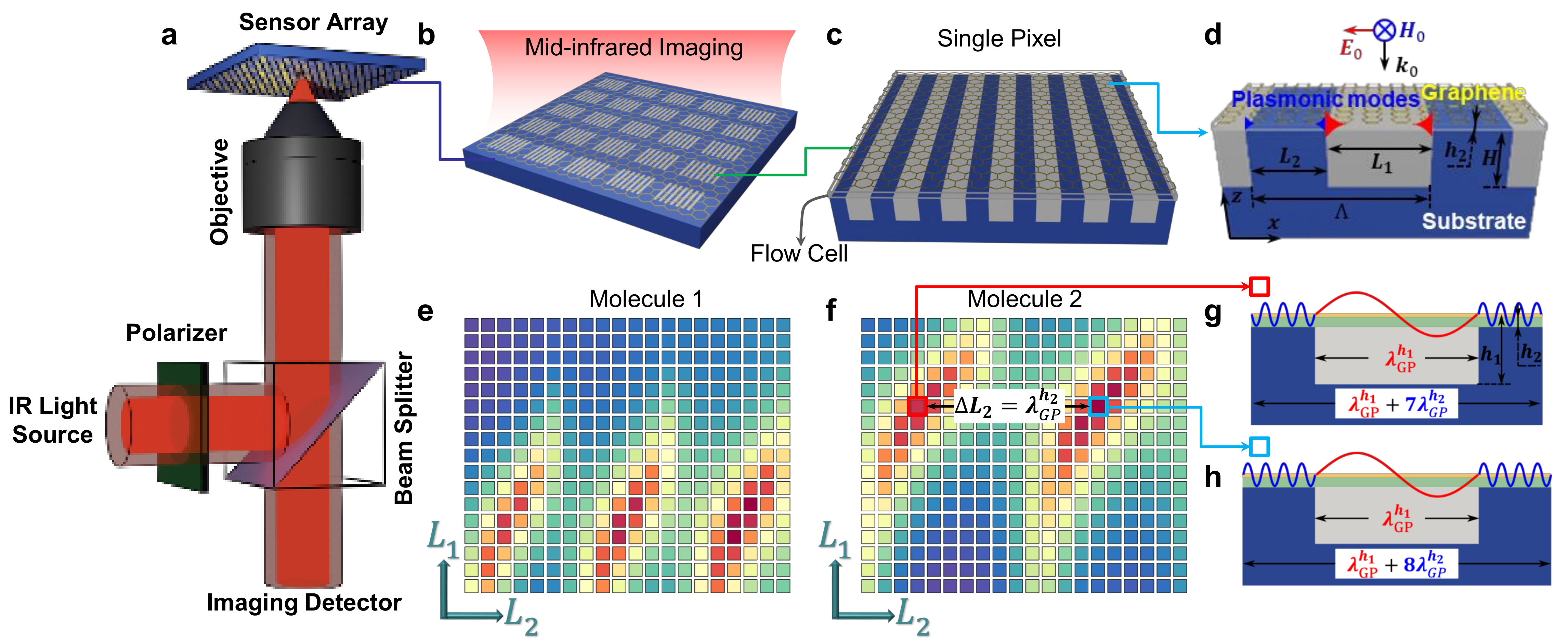}
%\par\end{centering}
\caption{\label{fig:configurations} Molecular refractive index detection with digital sensor using suspended graphene plasmon cavities. (a) Schematic illustration of the imaging-based MIR microscopy system. The monochromatic light generated by the light source focuses on a digital sensor after passing through a linear polariser, a beam splitter, and an objective. The incidence is a TM polarized light in terms of the grating surface, and normal to the surface of the sample. The reflected light is collected and imaged by an infrared camera. (b) The chessboard sensor is composed of a two-dimensional array of suspended graphene plasmon cavities with resonances that are highly sensitive to the refractive index of the analyte. (c) Each building block is composed of a graphene monolayer deposited on a metallic grating. A flow cell is used to circulate liquid samples and support the graphene layer. (d) Schematic of grating-assisted suspended graphene plasmon cavities. The incidence is transverse magnetic (TM) polarized in air, where $\Lambda$, $L_{1}$, $L_{2}$, and $H$ are the period, trench length, ridge length, height of the grating, and $h_{2}$ is the gap thickness. Thus, we have $h_{1}=h_{2}+H$. The gap and the trench can be filled with dielectric media.  (e-f) The captured absorption patterns for two different molecules. In this works,  we choose $L_1$ and $L_2$ as two variables of the array in the horizontal and vertical directions. It is worth noting that any one pair of the parameters (such as $L_{1}$, $L_{2}$, $h_{2}$, $H$ and $E_{F}$) of the graphene plasmon cavity system could be used as the varying parameters of the array. The material in the flow cell has refractive index $n$. (g-h) Sketches of two resonance modes for two structures with different ridge lengths. The difference of the ridge lengths equals one wavelength of GPs in the gap, which can be calculated using the height-dependent dispersion relation.}
\end{figure*}

%\textbf{\textcolor{red}{Target}}
 In this work, we report on a broadband tunable sensing technique based on suspended graphene plasmon cavities (Fig. \ref{fig:configurations}) and demonstrate its capability for enhancing light-matter interactions, detecting the real part of the refractive index of the analyte, and characterising both types and concentrations of the various molecules. Our design exploits the standing wave properties of plasmonic resonance in suspended graphene cavities,  which is driven by two horizontal Fabry-P\'{e}rot cavities and one vertical Fabry-P\'{e}rot cavity (Fig. \ref{fig:configurations}c and d). In particular, we implement a two-dimensional array of suspended graphene plasmon cavities, which consist of a continuous graphene monolayer on top of a metallic grating with a gap. The GPs excited due to the scattering from the edges of the cavities are horizontally reflected back and forward in the gap and in the trench to form two horizontal cavities (Fig. \ref{fig:configurations}g and h). The resonance is highly sensitive to the real part of the refractive index of the analyte and and we expect it to be visible in the reflection (absorption) pattern of the system due to the dissipation. Therefore, this configuration allows us to establish a unique relation between the real part of the refractive index of the analyte and the spatial absorption pattern (Fig. \ref{fig:comparassion}, and Fig. \ref{fig:configurations}e and f). By translating these spatial absorption maps for different molecules into their refractive indices, we demonstrate that this technique is suitable for refractive index measurement, chemical identification, and concentration analysis. The proposed technique offers several advantages in molecular sensing over traditional nanophotonic systems. These include the ability to measure refractive indices, the ultra-broadband measurable spectral range, the tiny volume of the analyte required, and the deeply subwavelength dimensions of the elements.

\section{Results}

%\textbf{\textcolor{red}{Theory of the suspended graphene plasmon cavities}}
Plasmonic waves in graphene have attracted numerous investigations, because of their extreme field confinement. However, the efficient excitation of such plasmonic waves is still very challenging because there is a large momentum mismatch between the incident light and GPs. In this work the GPS are excited efficiently and the extremely high field enhancement and extraordinary compression of GPs occur simultaneously, thanks to the combination of a shallow cavity and a deep cavity in the same configuration: the shallow one is above the ridge with length $L_{2}$ and height $h_{2}$, and the deep one is in the trench with length $L_{1}$ and height $h_{1}$ (Fig. \ref{fig:configurations}d). In each cavity, the scattering by sharp ridges is used to generate a broad spectrum of wavevectors to compensate for the momentum mismatch. The forward and backward launched plasmonic waves constructively interfere in the cavities to form standing waves (Fig. \ref{fig:configurations}g and h). Thus, when the cavity length $L_{\mathrm{j}}$ satisfies the Fabry-P\'{e}rot equation (where $j=1, 2$ represents the cavities in the trench and above the ridge, respectively), the GPs may be excited. A Fabry-P\'{e}rot equation reads
\begin{equation}
    \delta\phi_{\mathrm{j}}+\Re\{k_{\mathrm{GP}}^{h_{\mathrm{j}}}(\lambda_{0})\} L_{\mathrm{j}}=m\pi,\ m=0,1,2,3... ,\ \mathrm{j}=1 \mathrm{\ and\ } 2,
\end{equation}
where $\delta\phi_{\mathrm{j}}$ is the phase shift, $k_{\mathrm{GP}}^{h_{\mathrm{j}}}$ denotes the wavenumber of the GPs, and the integer $m$ denotes the resonance mode order. The cavity height-dependent dispersion relation of GPs derived in reference \citenum{xiao2018theoretical} is used in this work. The real part of the wavenumber of GPs determines the effective plasmon wavelength, $\lambda_{\mathrm{GP}}^{h} = 2\pi/\Re\{k_{\mathrm{GP}}^{h}\}$, where $h$ denotes the cavity height and could be $h_{1}$ or $h_{2}$ in this work. To study the properties of GPs, the effective refractive index (ERI 
or modal) of the GP, $N_{\mathrm{eff}}^{h} = k_{\mathrm{GP}}^{h}/k_{0}$ (where $k_{0}$ is the wavenumber in free space), is introduced. Taking a phase shift of $0$, we get $L_{\mathrm{j}}=m \lambda_{\mathrm{GP}}^{h_\mathrm{j}}/2$, which we find to be a good approximation for our results. It should be noted that the phase shift is frequency-dependent. Fig. \ref{fig:configurations}g and h show the sketches of the standing waves of GP resonances supported in this system. While those waves propagate in the graphene layer, the optical energy is dissipated thorugh Ohmic losses.

\begin{figure}[htbp]
%\begin{centering}
\includegraphics[width=0.8\linewidth]{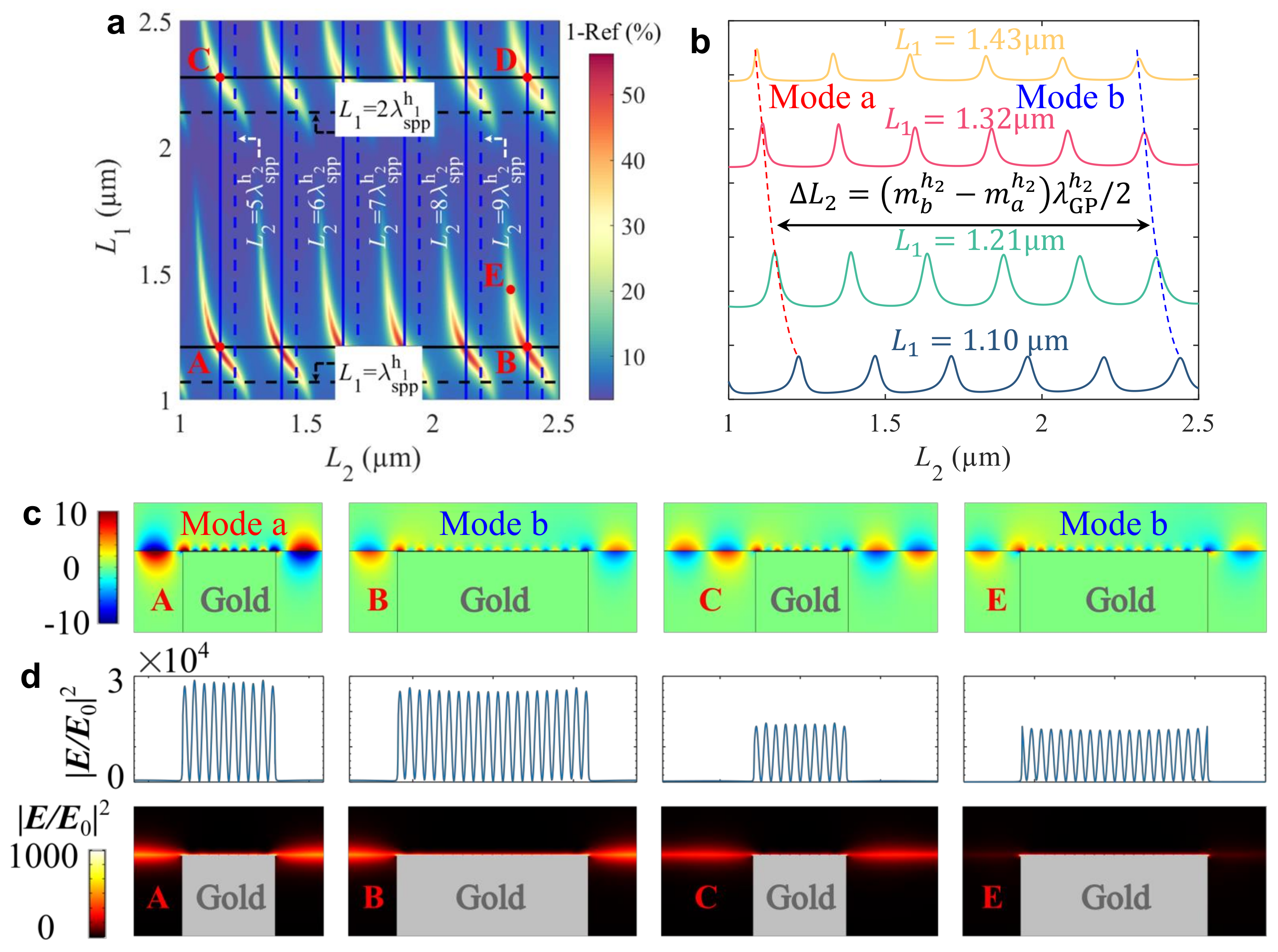}
%\par\end{centering}
\caption{\label{fig:workingprinciple} Resonance generation mechanisms in suspended graphene plasmon cavities. (a) Absorption of the system ($1-Ref$, where $Ref$ denotes the reflection) for the gold grating plus graphene structure versus the trench length ($L_{1}$) and the ridge length ($L_{2}$), when the gap thickness is $5$ nm. The normal incidence is at a wavelength $\lambda_{0} = 12$ \textmu m, where the trench and the gap are filled with air, and $E_{\mathrm{F}} = 0.64$ eV. The dashed lines indicate a fit to a Fabry-P\'{e}rot model for a phase shift of $0$. The black solid lines indicate a fit to a Fabry-P\'{e}rot model for a phase shift of $-0.27\pi$, while the blue solid ones indicate a fit to a Fabry-P\'{e}rot model for a phase shift of $0.54\pi$. (b) Line cuts of the absorption in (a) for four trench lengths ($L_1$) indicated by the labels. The orders for mode a and mode b are $ m_a^{h_2}=10$ and $ m_b^{h_2}=20$, respectively. (c) The component $E_{z}$ of the electric field is shown for the selected points marked by A, B, C, and E in (a). The corresponding parameters are as follows: A, $L_{1}=1.21$ \textmu m and $L_{2}=1.15$ \textmu m; B, $L_{1}=1.21$ \textmu m and $L_{2}=2.365$ \textmu m; C, $L_{1}=2.28$ \textmu m and $L_{2}=1.15$ \textmu m; E, $L_{1}=1.45$ \textmu m and $L_{2}=2.310$ \textmu m. The electric field is normalized to the incident field amplitude $E_0$ . (d) The electric near-field with high-intensity enhancement ($|E/E_0|^{2}$) is strongly compressed in the cavity above the ridge. Here, $\lambda_{0}/h_{2}=2400$. Distribution of the intensity along a line positioned in the middle of the gap is given as well.}
\end{figure}

% \section{Results}
%Configurations-
%\textbf{\textcolor{red}{Fundamental building block}}
Based on the Fabry-P\'{e}rot equation and the dispersion relation of GPs, we can establish a theory for the fundamental building block of our sensor, as shown in Fig. \ref{fig:configurations}c and d. The system is composed of a graphene monolayer deposited on a gold grating. We assume that there is a small gap (with a thickness of $h_{2}$) between the two. The gap and the trench are filled with the analyte and dielectric materials, respectively. Without loss of generality, we shall take the dielectric in the trench to be air. More realistic configurations will be discussed later. To investigate our system, we begin by calculating the absorption of the system as a function of trench length ($L_{1}$) and ridge length ($L_{2}$) for a fixed gap thickness $h =5$ nm and a fixed incident wavelength $ \lambda_0 = 12 $ \textmu m. According to the cavity height-dependent dispersion of GPs \cite{xiao2018theoretical}, the corresponding ERIs ($N_{\mathrm{eff}}^{h}$) are $49.3015+0.3239i$ and $11.2314+0.1315i$ with $h =5$ nm and $h >1$ \textmu m, respectively. The absorption versus the trench length and the ridge length is shown in Fig. \ref{fig:workingprinciple}a. The strong absorption peaks are due to the excitation of the GPs, which can be confirmed by near-field plots in Fig. \ref{fig:workingprinciple}c.
The component $E_{z}$ is antisymmetric as a consequence of the electromagnetic field boundary conditions and the symmetry of the system.

In earlier works \cite{li2017graphene,xiao2018theoretical}, the graphene monolayer and the grating were in physical contact with one another. However, due to the existence of the small gap in this study, there exists a very strong field in the parts of the graphene which are located above the trench of the gold grating. Consequently, there is very strong absorption in these locations. In Fig. \ref{fig:workingprinciple}a, the local maximum values of the graphene absorption correspond to the excitation of standing waves in both cavities. The predictions of those local peak positions are also given using the proposed theory, as the crossing points of the black dashed lines (for the cavity in the trench) and the blue dashed lines (for the cavity above the ridge) shown in Fig. \ref{fig:workingprinciple}a. Although the reasonable agreement between the predictions and the numerical results confirms the validity of the proposed model, there is an apparent shift between the predictions and the real local peak positions. This discrepancy is due to the coupling between the graphene plasmonic modes and the dipole mode of the trench, which can be observed clearly in the near-field plots shown in Fig. \ref{fig:workingprinciple}c. Therefore, we need to modify the Fabry-P\'{e}rot model by optimizing the phase shift $\delta \phi_{\mathrm{j}}$ to a value different from $0$ to compensate for the phase difference (which comes from the coupling between the GP mode and the trench dipole mode), as the figure caption shows. It should be noted that the phase shifts $\delta \phi_{\mathrm{j}}$ strongly depend on the gap thickness, the permittivities of the materials in the system, and the properties of the graphene layer. Using the modified phase shifts, we demonstrate excellent predictions as shown by the crossing points between the black solid lines (for the cavity in the trench) and the blue solid lines (the cavity is above the ridge) in Fig. \ref{fig:workingprinciple}a.

Another interesting feature  is that strong absorption persists when the conditions for establishing standing waves in those two cavities are not fully satisfied individually. In addition, the shape of the strong absorption pattern in Fig. \ref{fig:workingprinciple}a looks like a "Sigmoid function" with a rotation of 90 degrees. This is because plasmonic modes in both cavities can still be excited, as long as their phases compensate each other at the edges. This is confirmed by the near-field plot in Fig. \ref{fig:workingprinciple}c for the point marked as point E in Fig. \ref{fig:workingprinciple}a. The absorption is strongest when the Fabry-P\'{e}rot resonance is fully satisfied in each cavity.  It is worth noting that, if we choose several trench lengths, as shown in Fig. \ref{fig:workingprinciple}b, the difference in the position of the adjacent peaks equals the effective wavelength of GPs in the shallow cavities ($\lambda_{\mathrm{GP}}^{h_{2}}$), which can be used to retrieve the refractive index of the material in the gap.

%\textbf{\textcolor{red}{Enhancement and compression}}
One of the main appeals of such a suspended graphene cavity system is that the extremely high field intensity is extraordinarily compressed inside the gap above the ridge. Previous work \cite{xiao2018theoretical} has shown that the electromagnetic field can be strongly trapped in a very shallow cavity (the compression factor is about $\sim400$) and that high intensity enhancement ($\sim4000$) of graphene plasmon waves can be realized under certain conditions. However, these two phenomena did not appear in the same system. In this work, we achieve both the extremely high enhancement and the extraordinary compression in the same system, as shown in Fig. \ref{fig:workingprinciple}d. This is because: (i) the tiny gap above the ridge makes the extraordinary compression possible, and (ii) the optimized trench height could significantly enhance the conversion efficiency using a horizontal Fabry-P\'{e}rot cavity. Higher enhancement and even more extraordinary compression could be achieved by further optimizing the parameters of the system.

%\textbf{\textcolor{red}{Thickness dependence, tunability of the Fermi energy, variation of the refractive index in cavities}}
So far, we have confirmed that the resonance of the fundamental building block of the proposed sensor can be precisely predicted by the established theory based on the Fabry-P\'{e}rot cavities and the height-dependent dispersion relation of GPs. It was verified that the established theory remained valid when we changed the incident wavelength and the parameters of this configuration, such as the gap thickness, the cavity height, the Fermi energy of the graphene layer, and the materials filled in the gap and the trench. In the following, we will use the deeply subwavelength Fabry-P\'{e}rot cavities formed in the system for molecular refractive index measurement.

%\textbf{\textcolor{red}{Working principle of refractive index detection using a digital sensor}}
To illustrate the working principle of the refractive index measurement using our tiny Fabry-P\'{e}rot cavity, we first consider an ideal Fabry-P\'{e}rot cavity. Depending on the length of the cavity, a greater or smaller number of interference fringes inside the cavity are observed, which we usually call different orders of modes. According to the conventional Fabry-P\'{e}rot equation, the real part of the refractive index of the cavity core can be expressed as $n_c =\lambda_{0} /(2\Delta l)$, where $\Delta l$ is the difference of the cavity length for the adjacent modes. In a bulk system, the resonances are observed in the transmission of the cavities. This is also the core concept of our technique, though there is still a distinct difference. In our system, it is the GPs which form  deeply subwavelength Fabry-P\'{e}rot cavities. Consequently, $n_c$ in the above equation should be $\Re \{N_{\mathrm{eff}}^{h}\}$ in our technique.
 
To observe the resonances, we propose measuring the reflection (or absorption) pattern of a two-dimensional array of suspended graphene plasmon cavities, with two variables of the configuration in the horizontal and vertical directions (Fig. \ref{fig:configurations}). Although there are several choices for these two variables, we choose $L_{1}$ and $L_{2}$ here (Fig. \ref{fig:configurations}e and f), respectively, due to the convenience of the fabrication. The key points of the proposed technique are to extract the effective wavelength of GPs from the measured absorption pattern based on the Fabry-P\'{e}rot equation and then recalculate the refractive index of the analyte using the dispersion relation. Although there are several different approaches to realize this technique, in this demonstration (Fig. \ref{fig:workingprinciple}b), the value of $\lambda_{\mathrm{GP}}^{h_2}$ is extracted by calculating the separation of adjacent resonance peak positions via $\lambda_{\mathrm{GP}}^{h_2}=2\Delta L_2/(m_b^{h_2}- m_a^{h_2})=\lambda_0 /\Re \{N_{\mathrm{eff}}^{h_2}\}$, where $ m_b^{h_2}$ and $ m_a^{h_2}$ are  the orders of the modes excited, respectively, and $ \Delta L_2$ is the difference of the trench lengths for those two modes. Although the absolute values of $m_a^{h_2}$ and $ m_b^{h_2}$ are usually difficult to be measured in experiments, their difference would be easily obtained by counting the number of resonance peaks. It is also worth noting that, the "Sigmoid function" shape of the strong absorption pattern makes the resonances straightforward to find in the absorption pattern because even when the Fabry-P\'{e}rot resonance is not fully satisfied in each cavity, the modes 
can still be excited and thus be observable. The separation of the adjacent resonance peaks observed for different trench lengths is the same (Fig. \ref{fig:workingprinciple}b). Such a phenomenon is particularly attractive in the design of the sensor because it allows robust observation of the resonance modes and the adjacent resonance peaks in the captured image. The real part of the refractive index of the analyte could be numerically calculated based on the dispersion relation of GPs. When varying the incident wavelength, the refractive indices of the analyte at a certain spectral range can be obtained. Consequently, the type, even the concentration, of the molecule to be analyzed can be differentiated based on the retrieved refractive index.

\begin{figure}[htbp]
%\begin{centering}
\includegraphics[width=0.9\linewidth]{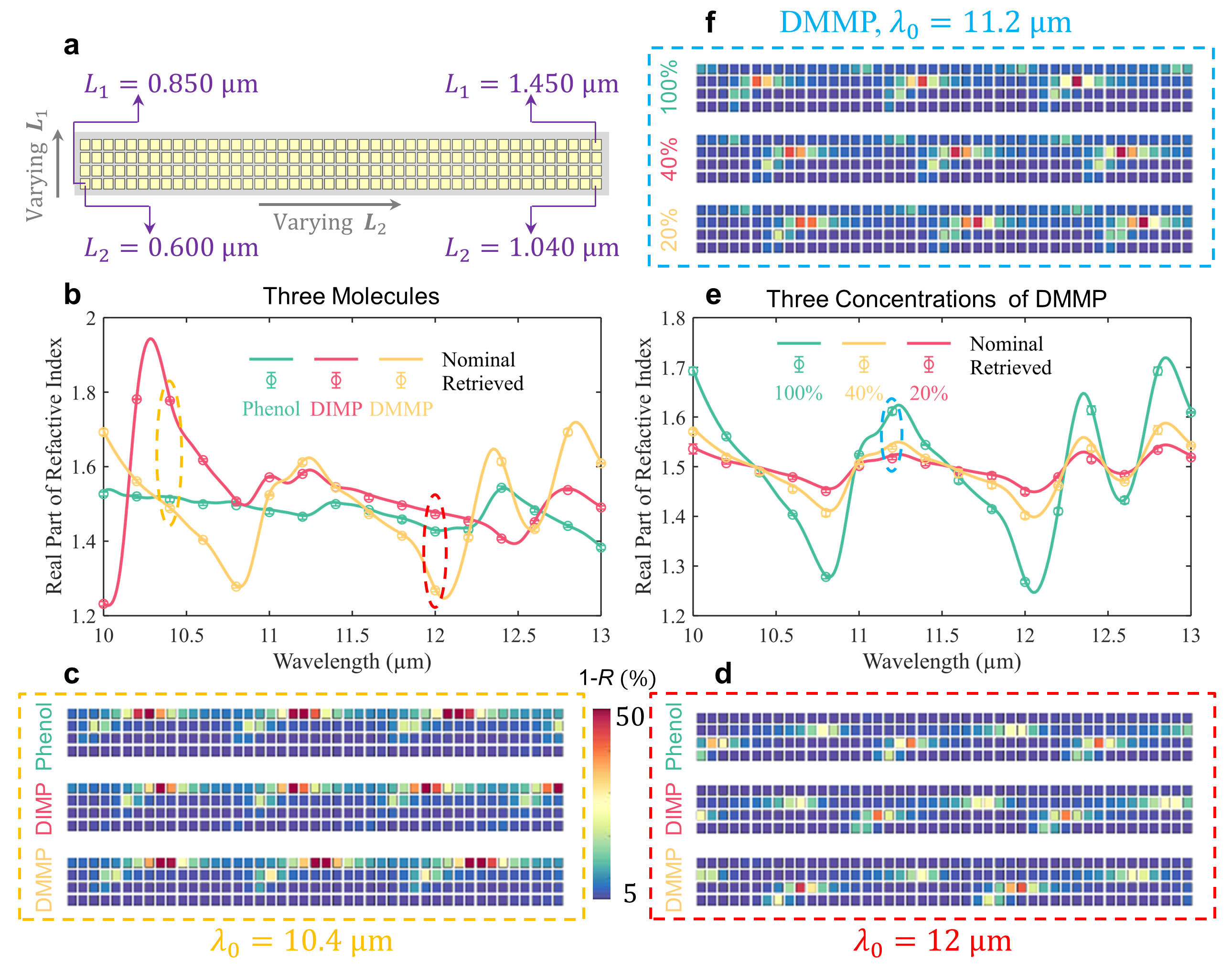}
%\par\end{centering}
\caption{\label{fig:Results} Molecular refractive index retrieval and spatial absorption mapping for chemical identification and concentration analysis. (a) a two-dimensional array of suspended graphene plasmon cavities with varying trench length and ridge length. The trench length in the unit cell varies linearly between $0.850$ \textmu m and $1.450$ \textmu m in $4$ steps in the vertical direction. The ridge length varies linearly between $0.600$ \textmu m and $1.040$ \textmu m in $45$ steps in the horizontal direction. (b) Nominal and retrieved refractive index of three different molecules. Absorption bar-codes of the pixelated sensor at $\lambda_{0} = 10.4 $ \textmu m (c) and $\lambda_{0} = 12 $ \textmu m (d) for three molecules. (e) Nominal and retrieved refractive index of three different concentrations. (f) Absorption bar-codes of the pixelated sensor at $\lambda_{0} = 11.2 $ \textmu m for three concentration of DMMP. }
\end{figure}

%\textbf{\textcolor{red}{Refractive index measurement for different molecules}}
We demonstrate the performance of the proposed sensor by showing its capability of detecting the real part of the refractive index for three different molecules. Without any loss of generality, we select the incident wavelength ranging from $10 $ \textmu m to $12 $ \textmu m. As shown in Fig. \ref{fig:Results}, a two-dimensional array of suspended graphene plasmon cavities with varying trench lengths and ridge lengths are used. It should be further noted that a flow cell is placed in the gap over the ridge and on top of the trench to circulate liquid samples and to be reusable (Fig. \ref{fig:configurations}c). The flow cell may also serve to support the graphene layer. The thickness of the flow cell is only $5$ nm, which means the volume of analyte required in the measurement is very small. The sensor is used to detect three different molecules, including phenol \cite{francescato2014graphene}, diisopropyl methylphosphonate \cite{querry1987optical} (DIMP) and dimethyl methylphosphonate \cite{querry1987optical} (DMMP). It should be noted that, in the simulation, only the real part of the reflective index of the analyte is used as proof of concept. The refractive index here is used to trigger the resonance of the system, and the energy is mainly absorbed by the graphene layer, rather than the analyte itself. This is particularly suitable for the measurement of the refractive index with negligible  imaginary part, in which case the approaches based on characteristic molecular absorption fingerprints would fail. As shown in Fig. \ref{fig:Results}b, the retrieved refractive indices of different molecules are in good agreement with the nominal ones, which confirms the validation of the proposed techniques.

%\textbf{\textcolor{red}{Refractive index measurement for different concentrations}}

Here, we also assess the capability of our sensor for concentration detection  (Fig. \ref{fig:Results}e). For techniques based on the SEIRA concept, when the detected absorption fingerprints of different molecules overlap, it is difficult to differentiate the types of analytes. In addition, for traditional nanophotonic sensors, the signal-to-noise ratio is relatively low. For these reasons, it is very challenging to determine the concentration of molecules using current nanophotonic systems. However, our approach is based on sensing the real part of the refractive index of the analyte. Therefore, this technique would be more suitable for the determination of both the types and concentrations of various molecules. Fig. \ref{fig:Results}b and e demonstrate that the proposed technique is suitable for refractive index measurement, chemical identification, and concentration analysis. Moreover, the refractive index retrieval at a particular wavelength in this technique relies only on a single measurement, which is an advantage in sensing. For example, we can measure the concentration based on one measurement. Such single measurement detection is beneficial to high-throughput measurement.

Until now, the method for the molecule detection is based on the retrieved refractive indices calculated from the spatial absorption maps. In fact, the barcode-like spatial absorption maps of the two-dimensional array can be used directly for a molecule or concentration sensing, when a database is prepared based on the proposed theory. As shown in Fig. \ref{fig:Results}, different barcode-like spatial absorption maps for different wavelengths, molecules (Fig. \ref{fig:Results}c and d) or concentrations (Fig. \ref{fig:Results}f) are easily observed. Thus, this direct digital imaging technique offers the potential for chemical identification and concentration analysis through pattern recognition based on a library of multiple molecular barcode signatures.
 
\begin{figure}[htbp]
%\begin{centering}
\includegraphics[width=0.9\linewidth]{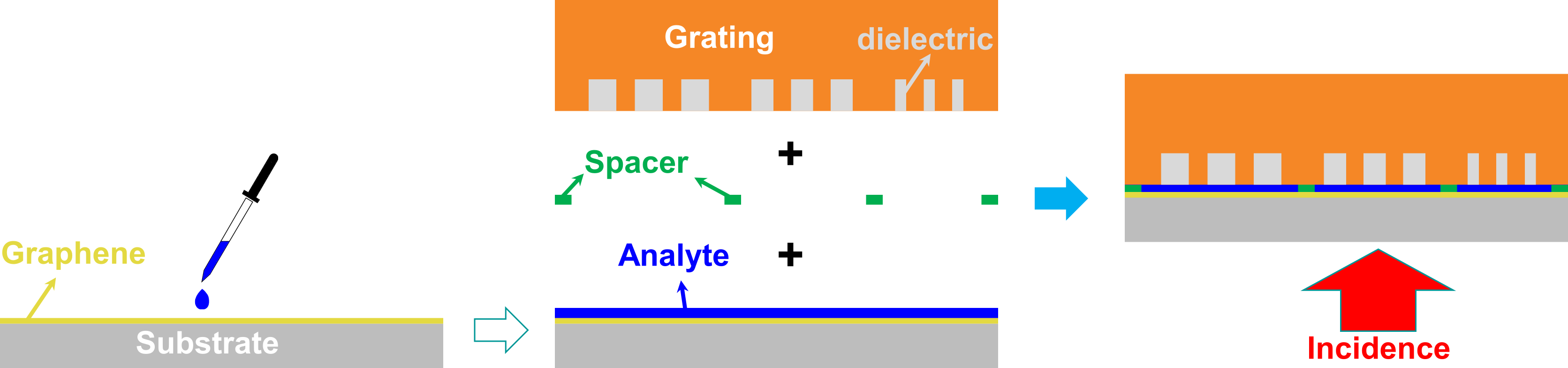}
%\par\end{centering}
\caption{\label{fig:Realistic} Schematic of a realistic configuration of a tunable digital sensor with suspended graphene plasmon cavities. The graphene is transferred on a substrate, such as Si substrate, which is transparent at the interested frequencies. The cavity in the metallic grating is filled with a dielectric medium. Put the liquid analyte on the substrate. Put the grating on the substrate. The gap with a desired thickness is achieved by varying a spacer between the substrate and the grating. The light is coming from the bottom substrate side.}
\end{figure}

Additionally, our scheme could provide an ultra-broad spectral measurement band, thanks to the super-broad band of GPs. When using resonance elements in traditional nanophotonic sensors, the resonance spectrum must be tuned to overlap with the molecular fingerprints. However, the tunability is very limited since it is achieved by varying the geometry or the optical properties (such as the permittivity) of the resonator. Hence, the spectral range measured by the traditional nanophotonic sensors is usually very narrow. In contrast, our technique is valid when the plasmons in the graphene layer can be excited efficiently, and thus its measurable spectral range could cover the mid- and far-IR and terahertz bands by carefully designing the two horizontal Fabry-P\'{e}rot cavities and the vertical Fabry-P\'{e}rot cavities. Finally, a more realistic configuration is suggested in Fig. \ref{fig:Realistic}. 

\section*{Discussions}
We have designed a digital sensor using suspended graphene plasmon cavities, which could provide extremely high field enhancement and extraordinary field compression. To do this, we have established the theory of the suspended graphene plasmon cavities. In the proposed theory, the Fabry-P\'{e}rot equation is used to predict the conditions for the GP resonance in two types of cavities (in the trench and above the ridge), and the dispersion relation of GPs in a general multilayer system is used to calculate the GP wavelength at different values of the cavity height. The excellent agreement between the predictions from this proposed theory and the numerical results confirms the validity of the proposed model. We also have achieved the extremely high field enhancement and extraordinary compression of GPs simultaneously, thanks to the combination of the shallow cavity and the deep cavity in the same configuration. Based on the excellent optical properties of such suspended graphene plasmon cavities, we have proposed the working principle of a digital sensor that allows the complex refractive index of an analyte to be retrieved. To demonstrate the performance of such a sensor, different molecules and different concentrations of DMMP have been (theoretically) detected. Given the capabilities of the refractive index measurement, the ultra-broadband measurable spectral range, the high sensitivity, and the small volume of the analyte required, the proposed sensor could serve as an ideal platform for applications such as chemical sensing, thermal imaging, heat scavenging, security and materials inspection, and molecule sensing \cite{germain2009optical, long2013recent,fan2008sensitive}. Furthermore, the tunability of the Fermi energy and geometric parameters of the cavities makes the design of this system extremely versatile. We also expect the investigated system using the combination of the shallow and deep cavities to open up numerous potential applications in photodetection, nonlinear optics, and integrated optics.

%\section*{Methods}
\section*{Materials and methods}
%\section*{Numerical Calculations}
\label{numbriecalCal}

\textbf{\textcolor{black}{Height-dependent dispersion relationship of GPs.}} In the previous study \cite{xiao2018theoretical}, we derived a dispersion relation of GPs in a general multilayer system, which can be described as a I-G-II-III system (where I, II and III represent metal or dielectric materials, and G represents a two-dimensional material). The corresponding optical properties are given by the relative permittivities $\epsilon_{i}$ where $i = 1, 2, 3$ represent media I, II and III, respectively, when only non-magnetic materials are considered. To describe the property of the interface G, a complex surface conductivity $\sigma_{g}$ is introduced. To obtain the dispersion relation of GPs in this system, we obtained the general TM solution. By considering the boundary conditions, the dispersion relation of the plasmon waves could be given by an implicit expression as follows \cite{xiao2018theoretical}
\begin{equation} \label{eq:dispersion}
    e^{-2k_{2}h}=\frac{k_{2}/\epsilon_{2}\delta+k_{1}/\epsilon_{1}}{k_{2}/\epsilon_{2}\delta-k_{1}/\epsilon_{1}}\frac{k_{2}/\epsilon_{2}+k_{3}/\epsilon_{3}}{k_{2}/\epsilon_{2}-k_{3}/\epsilon_{3}},
\end{equation}
where
\begin{equation}
\delta=1+i\sigma_{g} k_{1} /(\omega \epsilon_{0} \epsilon_{1}).
\end{equation}
Here $h$ denotes the thickness of slab II, $k_{i}^{2}=(k_{\mathrm{GP}}^{h})^2-k_{0}^{2}\epsilon_{i}$ denote the respective wavevector perpendicular to the interfaces in the three materials, and $\omega$ is the angular frequency of the wavelength of interest.

\textbf{\textcolor{black}{Numerical simulation.}} Finite-element method simulations were carried out using COMSOL Multiphysics to obtain the absorption properties and field distributions of this system. In this paper, the incident source is TM polarized and incident from the air side, normal to the surface, and the substrate material is gold. To simplify the analysis, the substrate is infinitely thick below the grating to switch off the transmission channel, and the dielectric material in the gap and trench is set as air. The details of the properties of graphene and gold may be found in the references \citenum{xiao2018theoretical} and \citenum{olmon2012optical}. To obtain a high absorption efficiency of the system, we have fixed the trench height ($H$) at $2 $ \textmu m, unless otherwise specified.

To show the capability of detecting the real part of refractive index, the analyte in the flow cell is assumed to have a real refractive index $n$, which is given by the real part of the refractive index of the molecule. To stress the ability of detecting the concentration ($co$) of the analyte, we approximate the refractive index as $n = 1.5 \times (1-co)+co \times n_{\mathrm{DMMP}}$ for different concentrations, where $n_{\mathrm{DMMP}}$ is the real part of refractive index of DMMP \cite{francescato2014graphene}.

\section{Conflicts of Interest}

There is no conflicts to declare.

\begin{acknowledgement}

X.X. is supported by Lee Family Scholars. S.M. acknowledges the Lee-Lucas Chair in Physics. V.G. acknowledges the Spanish Ministerio de Economia y Competitividad for financial support through the grant NANOTOPO (FIS2017-91413-EXP) and also the Ministerio de Ciencia, Innovaci{\'o}n y Universidades through the grant MELODIA (PGC2018-095777-B-C21). The authors thank Dr. Daan M. Arroo and Ms. Marie Rider for comments on the manuscript.
\end{acknowledgement}

%\begin{suppinfo}

%This will usually read something like: ``Experimental procedures and
%characterization data for all new compounds. The class will
%automatically add a sentence pointing to the information on-line:

%\end{suppinfo}

%\section{Author contributions}
%X. X. designed research. X. X. performed research; X. X., S. M. and V. G. analyzed data; X. X., S. M. and V. G. wrote the paper.

%\bibliographystyle{unsrtnat}
\bibliography{references}

\end{document}